# Mendeley Reader Counts for US Computer Science Conference Papers and Journal articles[1]

Mike Thelwall, Statistical Cybermetrics Research Group, University of Wolverhampton, UK.

Although bibliometrics are normally applied to journal articles when used to support research evaluations, conference papers are at least as important in fast-moving computing-related fields. It is therefore important to assess the relative advantages of citations and altmetrics for computing conference papers to make an informed decision about which, if any, to use. This paper compares Scopus citations with Mendeley reader counts for conference papers and journal articles that were published between 1996 and 2018 in 11 computing fields and had at least one US author. The data showed high correlations between Scopus citation counts and Mendeley reader counts in all fields and most years, but with few Mendeley readers for older conference papers and few Scopus citations for new conference papers and journal articles. The results therefore suggest that Mendeley reader counts have a substantial advantage over citation counts for recently-published conference papers due to their greater speed, but are unsuitable for older conference papers.

**Keywords**: Altmetrics; Mendeley; Scientometrics; Computer Science; Computing; Conference papers

## 1  Introduction

Altmetrics, social media indicators for the impact of academic research derived from the web (Priem, Taraborelli, Groth, & Neylon, 2010), are now widely available to help assess academic outputs. Altmetric.com, for example, collects a range of data about online mentions of academic documents, supplying it to journal publishers to display in article pages, to institutions to help them analyse their work and to researchers to track the impact of their publications (Adie & Roe, 2013; Liu & Adie, 2013). Many studies have investigated the extent to which altmetrics can be helpful for impact evaluations, including a few showing that early altmetric scores correlate with longer term citation counts (Eysenbach, 2011; Thelwall & Nevill, 2018). A limitation of almost all prior research is that it has focused on altmetrics for refereed journal articles, whereas monographs, conference papers or other outputs can be more important in some fields. This article assesses the value of one key altmetric, Mendeley reader counts, for conference papers. Although one small scale investigation has previously investigated this (Aduku, Thelwall, & Kousha, 2017), a comprehensive evaluation is needed.

Conference papers are known to be as important as journal articles in some areas of computer science, at least in terms of attracting as many citations (Freyne, Coyle, Smyth, & Cunningham, 2010; Goodrum, McCain, Lawrence, & Giles, 2001; Vrettas & Sanderson, 2015) and may be more important for computer science than any other field (Lisée, Larivière, & Archambault, 2008). Software Engineering journal articles indexed in Scopus have been shown to be more cited on average (arithmetic mean) than conference papers in the long term (Garousi & Fernandes, 2017). An investigation of Chinese computer science research has shown that the relative citation impact of journal articles and conference papers varies

---

substantially by field (Qian, Rong, Jiang, Tang, & Xiong, 2017), however, and so the software engineering results should not be generalised and any comparison in computer science must cover all fields to give general results. One general pattern is that conference papers become obsolete (stop attracting new citations) much sooner than do journal articles (Lisée, Larivière, & Archambault, 2008), perhaps because conferences focus more on fast-moving topics. Conference publishing can lead to double counting for citations if a conference is indexed by, for example, the Web of Science, and articles published based on these papers are also indexed (Bar-Ilan, 2010; González-Albo & Bordons, 2011). Such follow-up publications are the exception in computer science, however (Wainer & Valle, 2013).

Mendeley is a social reference sharing site owned by Elsevier but formerly independent. It is free to join and allows researchers to create their own libraries of papers that they plan to cite, supporting the creation of reference lists from them (Gunn, 2013). It also has academic social network features (Jeng, He, & Jiang, 2015). Here, the number of users that have registered a document in the site is its *Mendeley reader count*. Although these users have not necessarily read the document, most users add documents that they have read or intend to read (Mohammadi, Thelwall, & Kousha, 2016) and so "reader count" is reasonable terminology. Although altmetrics were originally believed to reflect non-scholarly impacts of research, Mendeley users are predominantly academics or doctoral students, with a small proportion of other students. In consequence, Mendeley reader counts correlate moderately or strongly with citation counts in most fields (Costas, Zahedi, & Wouters, 2015; Haustein, Larivière, Thelwall, Amyot, & Peters, 2014; Thelwall, 2017a; Zahedi & Haustein, 2018) and can be thought of as scholarly impact indicators (Thelwall, 2018), with an element of educational impact (Thelwall, 2017c). Reader counts seem to be one of the best known altmetrics (Aung, Zheng, Erdt, Aw, Sin, & Theng, 2019). Mendeley readers may not be common for other types of document, however, including preprints (Bar-Ilan, 2014). Their value is as early impact indicators because they appear about a year before citations (Pooladian & Borrego, 2016; Thelwall, 2017b; Zahedi, Costas, & Wouters, 2017), typically starting with the publication month of an article (Maflahi & Thelwall, 2018), allowing evaluations to be conducted more promptly (Kudlow, Cockerill, Toccalino, Dziadyk, Rutledge, Shachak, & Eysenbach, 2017; Thelwall, Kousha, Dinsmore, & Dolby, 2016).

The one published study of Mendeley readers for conference papers (Aduku, Thelwall, & Kousha, 2017) analysed Scopus journal articles and conference papers published in 2011 in two computing categories (Computer Science Applications; Computer Software) and two engineering categories (Building & Construction Engineering; and Industrial & Manufacturing Engineering). Conference papers in the two Engineering subjects and Computer Science Applications were rarely cited and rarely had any Mendeley readers. In contrast, Computer Software journal articles and conference papers were usually cited and with many Mendeley readers. There was also a strong Spearman correlation between the two for this category (journal articles: 0.572; conference papers: 0.473) (Aduku, Thelwall, & Kousha, 2017). This strong correlation, together with evidence of Mendeley use for Computer Software conference papers suggests that Mendeley may be useful for some computing conference papers, but perhaps not for all. Computer science journal articles are some of the least registered on Mendeley, however (Zahedi & van Eck, 2018).

Another reference manager, CiteULike (Emamy & Cameron, 2007; Sotudeh, Mazarei, & Mirzabeigi, 2015; Sotudeh & Mirzabeigi, 2015), has also been investigated for 1294 sampled computing-related conference papers from a conference support system, finding that the number of CiteULike readers (or bookmarks), associated with longer term CiteULike reader counts (Lee & Brusilovsky, 2019). Mendeley was not included, although it is more used than CiteULike in most fields for journal articles (Li, Thelwall, & Giustini, 2012; Thelwall, Haustein, Larivière, & Sugimoto, 2013). The reference manager Bibsonomy has not been investigated for computing. It has a small user base but a computing focus with a substantial minority of conference papers (Borrego & Fry, 2012). Connotea (Du, Chu, Gorman, & Siu, 2014) is also a free social reference sharing site.

The research goal of this article is to systematically evaluate Mendeley readership counts for conference papers over a long period in all areas of computing. The main restriction is to exclude papers with no authors from the USA. This step was made to focus on a country that is dominant in computer science and producing relatively high citation impact research. Conferences can sometimes be national and low quality, so a focus on the USA reduces the chance that these conferences could contaminate the results. The research questions are as follows.

- RQ1: In which publication years and fields are Mendeley readers more useful than citations for US computer science conference paper impact assessment?
- RQ2: In which publication years and fields are Mendeley readers more useful than citations for US computer science journal article impact assessment?
- RQ3: In which computer science fields are do Mendeley reader counts reflect a similar type of impact to citation counts?

## 2 Methods

### 2.1 Data

Elsevier's Scopus database was chosen as the source of the computer science conference papers and journal articles to investigate. Google Scholar indexes more computing citations (Franceschet, 2009; Martín-Martín, Orduna-Malea, Thelwall, & López-Cózar, 2018) but does not allow automatic harvesting of records by journal or conference, with the partial exception of the Publish or Perish software. Preliminary testing suggested that Scopus indexed more conferences than the Web of Science. The Scopus primarily journal-based classification scheme (https://www.elsevier.com/solutions/scopus/how-scopus-works/content, Source title list spreadsheet, ASJC tab) was used to organise the records by field. Although article clustering approaches and other classification schemes (e.g., ScienceMetrix) seem to be more internally coherent (Klavans & Boyack, 2017), the Scopus scheme is used for research evaluations and results based it are more transparent and easily reproducible than the alternatives. The two generic computer science categories, Computer Science (all) and Computer Science (misc) were not used since these do not correspond to fields.

All journal articles and conference papers published between 1996 and 2018 with at least one US author affiliation were downloaded during May 2019 using the Scopus API with queries like the following, one for each publication year (sent as a separate parameter). The code number at the start is the category code. For example, 1708 is Hardware and Architecture. These All Science Journal Classification (ASJC) codes can be found at the Elsevier URL above.

- SUBJMAIN(1708) AND DOCTYPE(ar) AND SRCTYPE(j) AND AFFILCOUNTRY("United States")
- SUBJMAIN(1708) AND DOCTYPE(cp) AND SRCTYPE(p) AND AFFILCOUNTRY("United States")

These queries produced 877,045 conference papers and 511,754 journal articles with at least one author from the USA 1996-2018.

For each article, the Mendeley API was queried via the free software Webometric Analyst (lexiurl.wlv.ac.uk) in May 2019 for the number of Mendeley readers. For papers or articles without a DOI, the query used the title, authors and publication year to get a set of potentially matching records from Mendeley (example queries are given in the Discussion). These were then filtered by Webometric Analyst to remove non-matching records. Following best practice, articles or papers with DOIs were also queried by DOI for additional matching records. When multiple records were found then they were combined to give the most complete results (Zahedi, Haustein, & Bowman, 2014)

## 2.2 Analysis

For RQ1 and RQ2, the usefulness of Mendeley readers in comparison to Scopus citations was assessed by identifying which is numerically the most common, in terms of the highest per-paper averages. Although this assesses quantity and not quality, correlation tests (RQ3) supplement the answers with information related to quality (as indicators of impact), as discussed below. Other factors being equal, an indicator derived from discrete data with many zeros is more useful if has higher average values. This is because there are likely to be fewer ties and so a better chance of differentiating between individual articles and more clearly differentiating between the average impacts of sets of articles. The trajectory of citation and reader counts will also be assessed visually to determine how soon counts approximate their final value, and therefore closely reflect the final or total citation/readership impact of documents.

Since sets of Mendeley reader counts (Thelwall & Wilson, 2016) and Scopus citation counts (de Solla Price, 1976) are highly skewed and close to lognormally distributed (Thelwall & Wilson, 2016; Thelwall, 2016a), the arithmetic mean is an unsuitable measure of central tendency (Fleming & Wallace, 1986; Limpert, Stahel, & Abbt, 2001). The geometric mean (Fairclough & Thelwall, 2015; Zitt, 2012) was used instead to assess average citation and reader counts. These were calculated separately for each field, year and document type (article or paper) because these three factors influence citation rates.

For RQ3, the extent to which Mendeley reader counts reflect a similar type of impact to citation counts was assessed only using correlation tests, as is standard for altmetrics (Sud & Thelwall, 2014). Positive correlations do not prove cause and effect relationships, although some Mendeley readers presumably use this reference manager to create citations, so there is a degree of cause-and-effect in the data. Nevertheless, most scientists don't use Mendeley (Van Noorden, 2014) so no overall causal connection can be claimed. In this absence, a positive correlation implies the existence of an underlying factor influencing both citation counts and Mendeley reader counts. Although there are no clear guidelines for interpreting the magnitude of correlations between citation counts and other indicators because of discrete data effects (Thelwall, 2016b) it is reasonable to interpret correlations around 0.5 or higher as evidence that citations and readers reflect very similar types of impact, especially when the average counts are low. This is reasonable because discrete data effects combined with low average numbers results in correlation coefficients that underestimate the strength of the underlying relationship (Thelwall, 2016b). Spearman correlations were used instead of Pearson correlations, again because of the skewing problem.

## 3   Results

The numbers of conference papers and journal articles (Fig 1) indexed by Scopus has increased reasonably steadily in most categories, although with some areas of decline, such as Computational Theory and Mathematics conference papers since 2009. Scopus indexes more conference papers than journal articles overall and Computer Networks and Communications is notable for having relatively many conference papers compared to journal articles indexed.

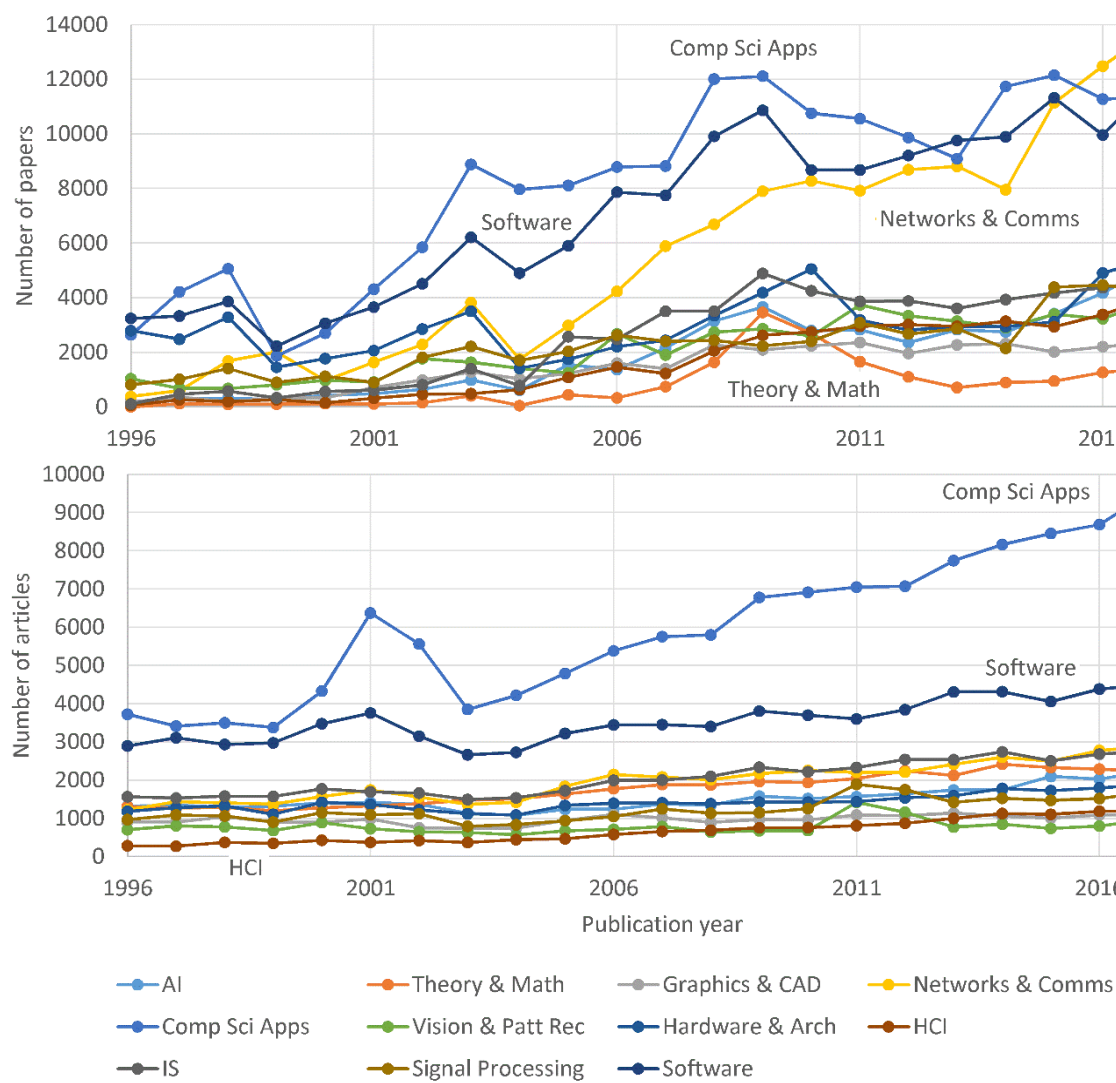

Figure 1. The number of US conference papers (top) and journal articles (bottom) by publication year and Scopus category. Individual fields can be identified in the versions of the graphs within Excel in the online supplementary materials.

Except for papers published in the year immediately preceding data collection, 2018, most conference papers and journal articles (Fig 2) have been cited.

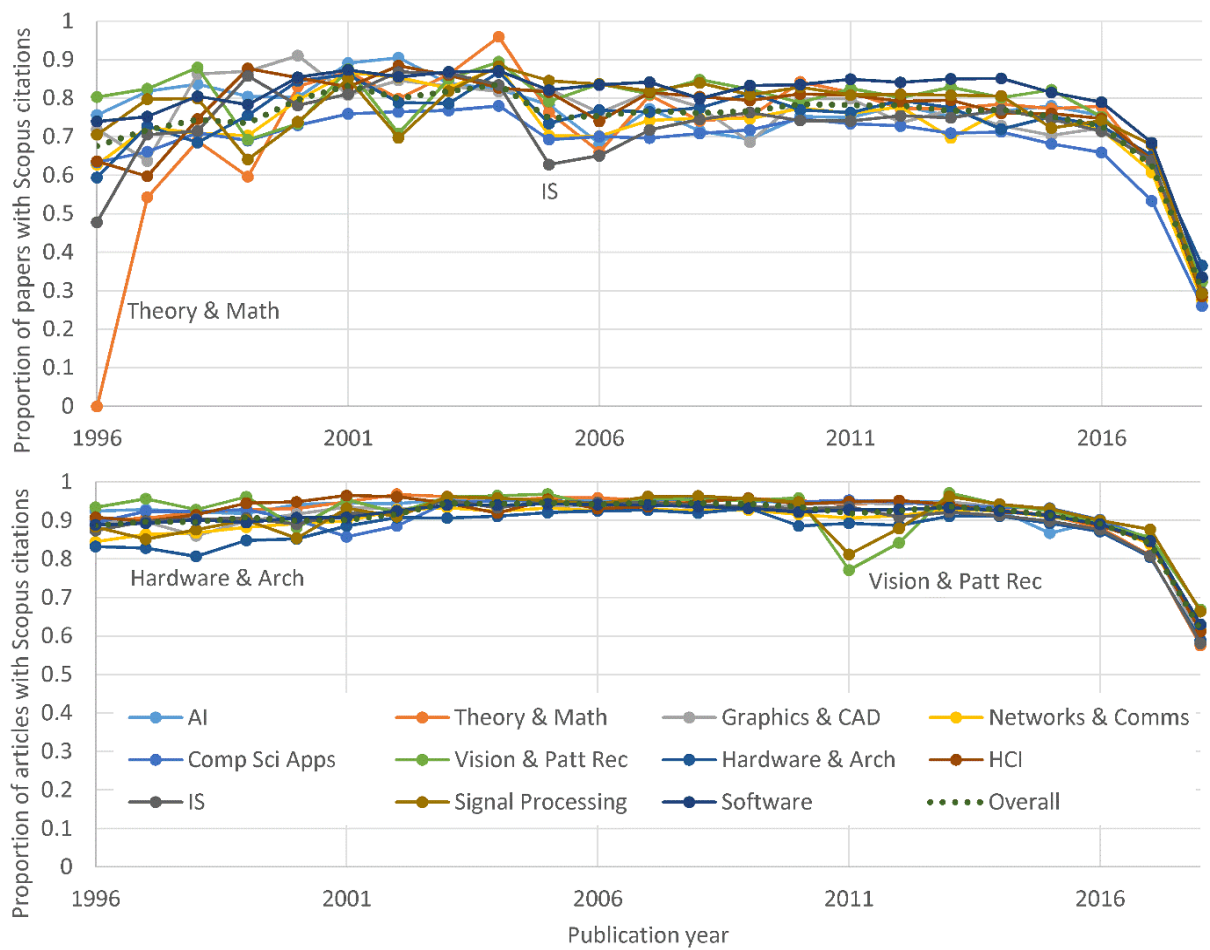

Figure 2. The proportion of US conference papers (top) and journal articles (bottom) with at least one Scopus citation by publication year and Scopus category.

The situation for Mendeley readers is quite different to that for Scopus citations. In all fields except one, most conference papers since 2006 have at least one Mendeley reader, but older conference papers are less likely to have Mendeley readers (Fig 3). For all years, a clear majority of journal articles have Mendeley readers (Figure 3, bottom). Thus, Mendeley users seem to ignore older conference papers much more than older journal articles. This is plausible if journal articles tend to have longer term significance than conference papers in the same field.

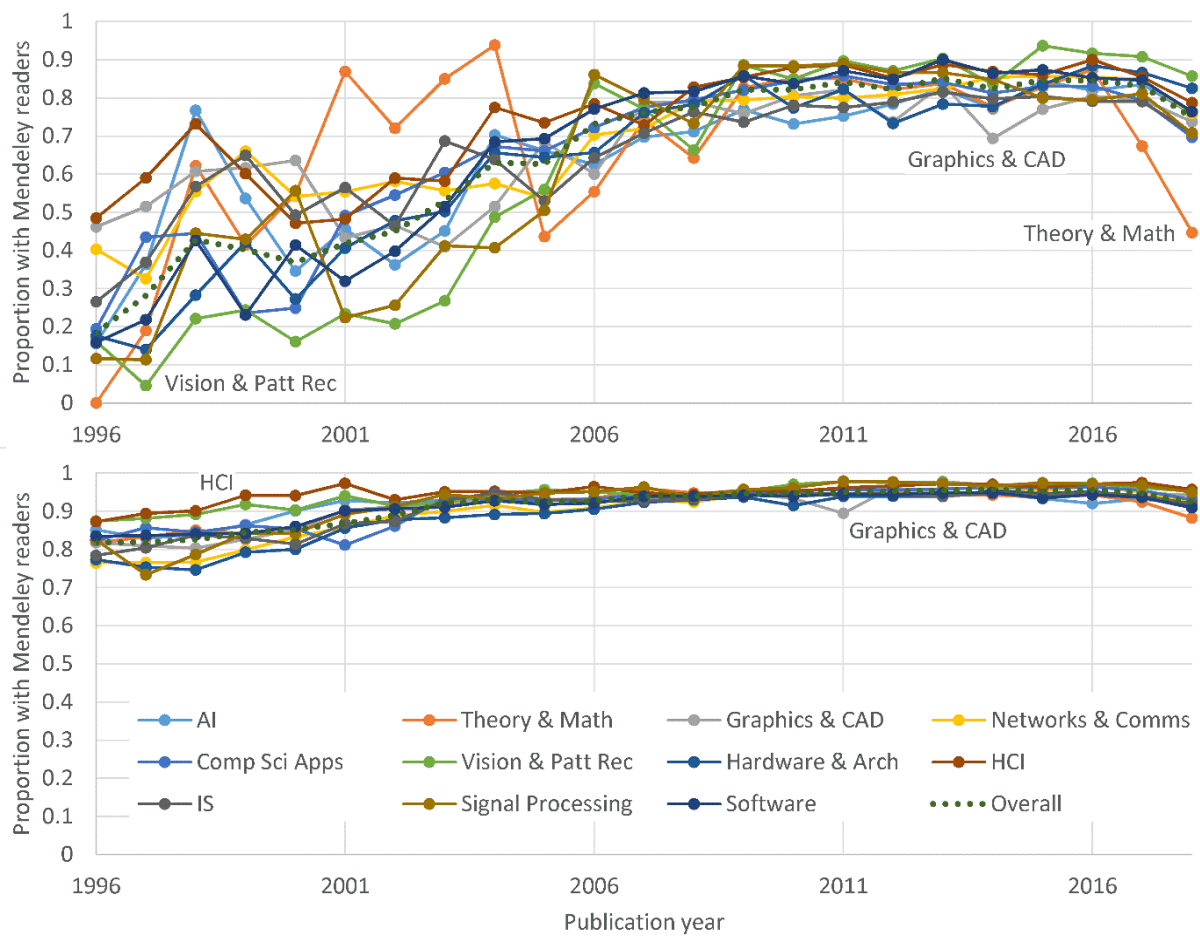

Figure 3. The proportion of US conference papers (top) and journal articles (bottom) with at least one Mendeley Reader (as returned by Mendeley API searches) by publication year and Scopus category.

Considering papers in reverse order (from newest to oldest), the average number of Scopus citations for conference papers (Fig 4) tends to increase from 2018 to 2011 and then stabilise. Higher values in some fields 1998-2004 may be due to not retrospectively indexing lower impact conferences, perhaps because they did not have online proceedings. For journal articles (Fig 4, bottom), the lower average citation counts before 2001 might be due to the relatively low total number of computing publications indexed until 2003, when Computer Science Applications started its rapid increase for conference papers and Computer Networks started its rapid increase for journal articles (Figure 1). Expanding category sizes can increase citation counts for recent articles because these are the most likely to be cited and there are relatively many citing articles compared to the number of cited articles.

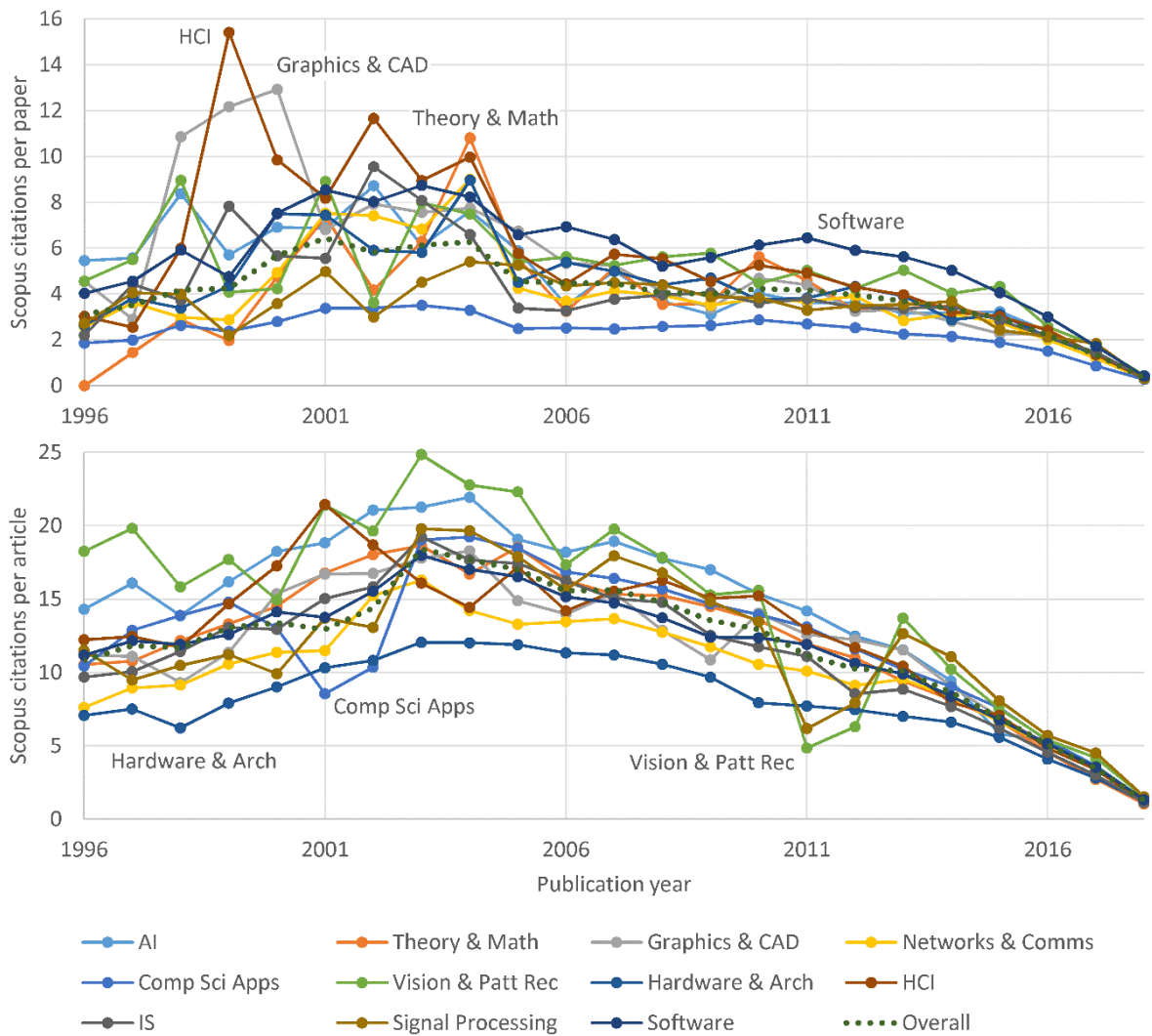

Figure 4. Average (geometric mean) Scopus citations for US conference papers (top) and journal articles (bottom) by publication year and Scopus category.

The average Mendeley reader counts for conference papers and journal articles (Fig 5) largely mimic the situation for the proportions of articles cited. Compared to Average Scopus citations, however (Fig 4), Mendeley reader counts tend to be higher than Scopus citation counts for both journal articles and conference papers published after 2006.

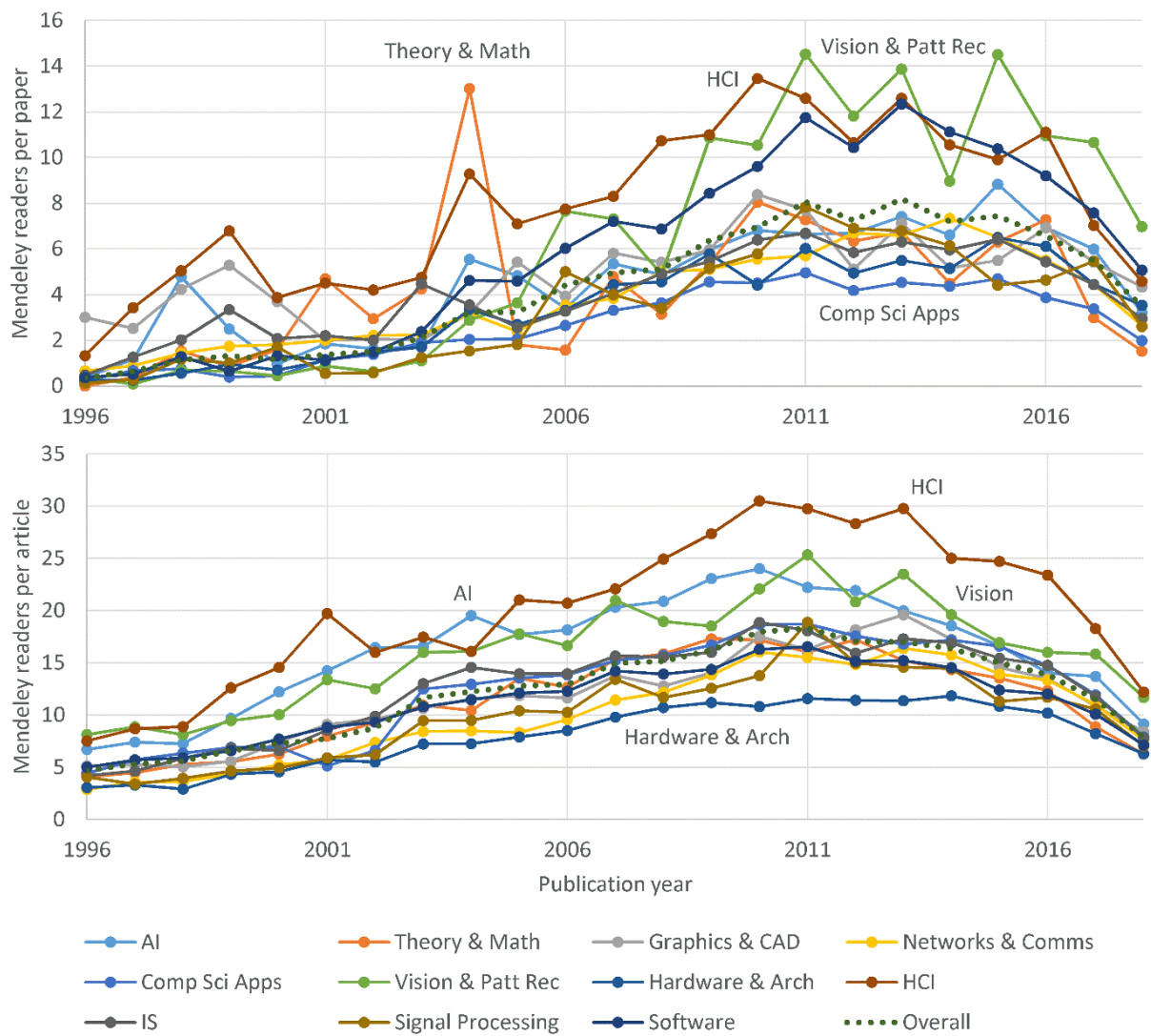

Figure 5. Average (geometric mean) Mendeley readers (as returned by Mendeley API searches) for US conference papers (top) and journal articles (bottom) by publication year and Scopus category.

Correlations between Mendeley reader counts and Scopus citation counts for conference papers are mostly moderate to strong 2006-2015, but weaker before 2016, presumably due to the many papers without Mendeley readers (Fig 6). Correlations between Mendeley reader counts and Scopus citation counts for conference papers are strong for all years, even the most recent year (2018) (Fig 6, bottom), with exceptions analysed in the Discussion. Computer science attracts citations to recently published journal articles relatively quickly to allow this high correlation, perhaps because of extensive and rapid conference publishing.

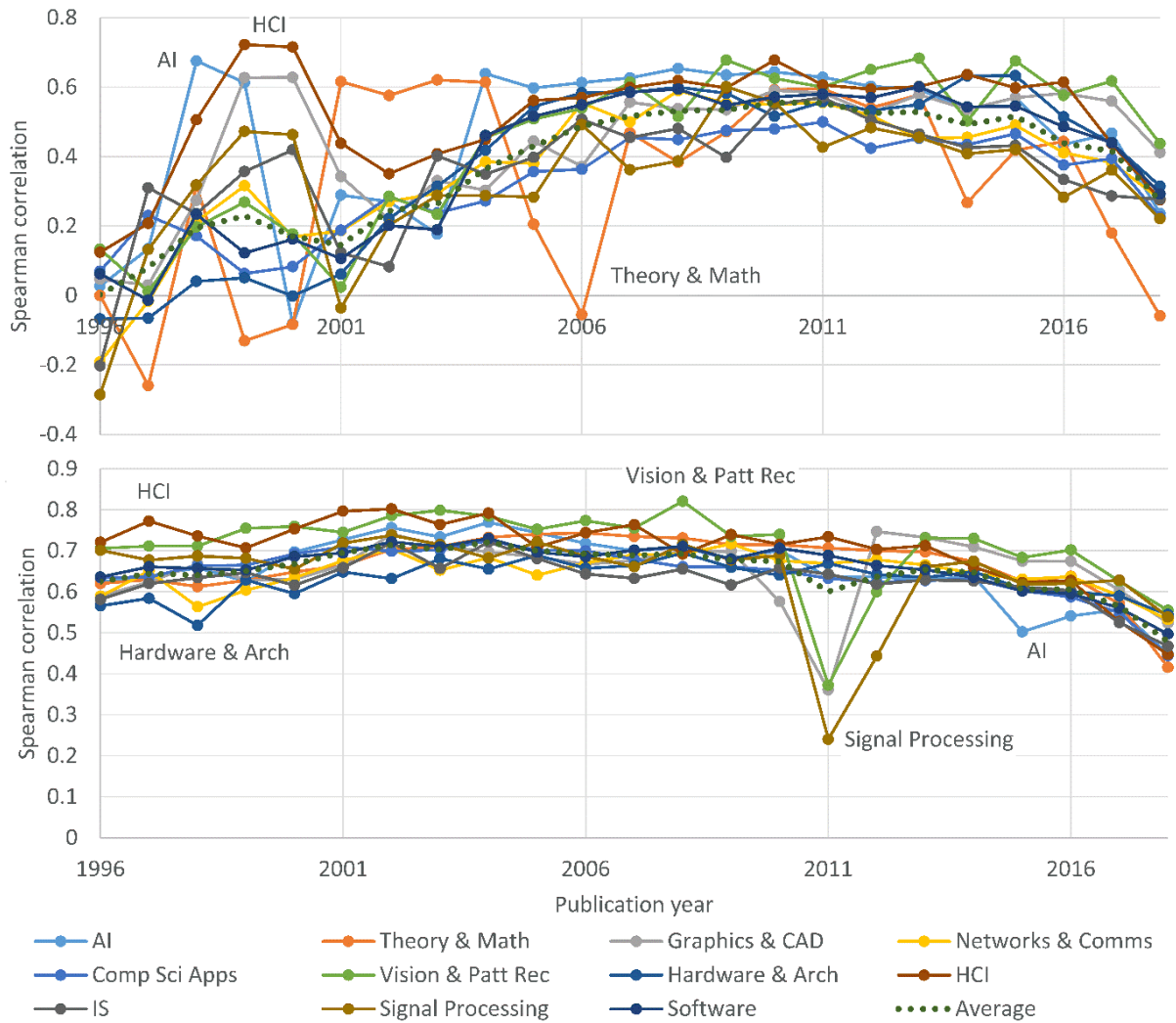

Figure 6. Spearman correlations between Mendeley readers (as returned by Mendeley API searches) and Scopus citation counts for US conference papers (top) and journal articles (bottom) by publication year and Scopus category.

## 4 Discussion

### 4.1 Anomalies

The conference paper correlations were low in 2006 for the category Computational Theory and Mathematics (Figure 6, top). The root cause was that in the low years, most papers in the conference *Empirical Methods in Natural Language Processing (EMLNP)* were not found in Mendeley. For example, the "Automatically assessing review helpfulness" EMNLP article from 2006 had 287 citations but was not found in Mendeley with the query:

- title:Automatically assessing review helpfulness AND author:Kim AND year:2006

Some other 2014 EMNLP articles were found, such as "Domain adaptation with structural correspondence learning" with 691 citations and 438 Mendeley readers, as found by the Mendeley query:

- title:Domain adaptation with structural correspondence learning AND author:Blitzer AND year:2006

In these years (2006 and 2014) Scopus had not indexed the DOIs of these papers and the title/author search often returned no hits for unknown reasons. The computational linguistics EMNLP conference had a substantial impact on the category because its papers were more cited than average for the field. Thus, the root causes are the combination of (a) a single relatively high citation conference, (b) Scopus not indexing paper DOIs and (c) the imperfect Mendeley search algorithm.

The journal article correlations were low in 2011 for three categories (Figure 6, bottom). One root cause was Scopus double-indexing conference papers as journal articles in this year and splitting their citations between the two versions. In Signal Processing and Computer Vision, there were 721 publications indexed as journal articles from one source (*Proceedings of the Annual International Conference of the IEEE Engineering in Medicine and Biology Society, EMBS*) and indexed as conference papers from another (*Conference proceedings : ... Annual International Conference of the IEEE Engineering in Medicine and Biology Society. IEEE Engineering in Medicine and Biology Society*). For example, "Modeling cortical source dynamics and interactions during seizure" had 4 citations linked with one version and 15 with the other, with the correct value presumably being 19.

In Computer Graphics and Computer-Aided Design, and also Computer Vision and Pattern Recognition, some articles in the *ACM Transactions on Graphic*s journal (a conference special issue) had two DOIs, one from the journal, and one from the host conference *Proceedings SIGGRAPH '11 ACM SIGGRAPH 2011*. Mendeley has sometimes picked a different DOI to Scopus for each article/paper but had merged articles so that a count of 0 would be returned if it had picked the conference DOI for the article/paper. For example, "Blended intrinsic maps" had the DOI 10.1145/2010324.1964974 in ACM ToG and in SIGGRAPH had the related DOI 10.1145/1964921.1964974.

## 4.2   Comparison with prior research

The results extend prior findings for two computing categories and one year (Aduku, Thelwall, & Kousha, 2017) by revealing universal patterns. The weak results previously found for Computer Science Applications (Aduku, Thelwall, & Kousha, 2017) are not typical for computer science generally because this category has the fewest citations and Mendeley readers for conference papers in most years, although it is average for journal articles. Computer Science Applications is the largest conference category for Computer Science. By far the biggest conference indexed in this category is the multidisciplinary *Proceedings of SPIE The International Society for Optical Engineering*, so the relatively low citations might be due to the incorporation of non-computer science papers from fields where conferences are less important.

The steeper initial slope and quicker flattening of the graph shapes for average Scopus citations to conference papers in contrast to journal articles (Figure 4) conflict with a decade-old finding that conference papers are cited more quickly (Lisée, Larivière, & Archambault, 2008), perhaps due to online first publishing. The results agree with previous evidence that conference papers become obsolete (no longer cited) more quickly (Lisée, Larivière, & Archambault, 2008). This result is confirmed here from the perspective of typical articles for the first time, since the geometric mean is used here.

Previous research has compared the importance of journal articles and conference papers in fields based on the total number of citations received (Freyne, Coyle, Smyth, & Cunningham, 2010; Goodrum, McCain, Lawrence, & Giles, 2001; Vrettas & Sanderson, 2015). Using the year 2006 for comparisons in Figure 4 (older years are unstable for conferences), on average, Scopus-indexed journal articles attract substantially more long-term citations than Scopus-indexed conference papers from the same field. Although this might be due to Scopus indexing lower quality conferences than journals, the restriction to US-authored articles makes this explanation unlikely.

Whilst there are differences between fields in average citation counts and reader counts for both journal articles and conference papers, echoing (Aduku, Thelwall, & Kousha, 2017) for two computing fields, there are universally moderate or high correlations and the differences are not large enough to suggest that Mendeley is substantially less useful for any field. The case with the weakest evidence to support its use is conference papers in Computer Science Applications. This is presumably due to the inclusion of multidisciplinary conferences, as discussed above.

## 5  Conclusions

The peak high correlations between Scopus citations and Mendeley readers for journal articles (above 0.6 for most years for all fields before 2010) and the moderate or high peak correlations for conference papers (above 0.4 for most years for all fields 2007-2015) suggest that Mendeley reader counts and Scopus citation counts probably reflect similar types of impact for conference papers (RQ3) and journal articles. The lower figures can be explained by smaller quantities of data (as discussed above). Combining this with the earlier appearance of Mendeley readers, it seems reasonable to use Mendeley readers as early citation impact indicators for conference papers in all areas of computing (RQ1), even in the year immediately following publication (e.g., for 2018 papers in the results here). Care should be taken with anomalies due to double-indexed conferences for some years, however.

Mendeley should be used cautiously for computing conference papers published before 2006, since the relative lack of Mendeley readers for these older articles and the lower correlations suggest that these reader counts are less reliable. For example, they may only cover papers used in education or classic papers that have not been converted into journal articles.

For journal articles (RQ2), Mendeley reader counts could reasonably be used as impact indicators for any of the years examined (1996-2018) based on high correlations with Scopus citations and geometric means that are not very low.

On the basis that, other factors being equal, higher average indictor values are the best evidence of usefulness, Mendeley reader counts are more useful than Scopus citation counts for computing conference papers immediately after publication and back as far as around 2006, when they have similar values. Similarly, for journal articles, Mendeley reader counts would be more useful than Scopus citations for publication dates after around 2001. In practice, since citations are probably more trusted than reader counts, it would be safer to use Mendeley readers for only the most recent three years after publication. After this, there would be a sufficiently wide citation window (Abramo, Cicero, & D'Angelo, 2011) to make Scopus citations reliable.